# Deterministic direct growth of WS$_2$ on CVD graphene arrays

*G. Piccinini[1,2], S. Forti[1], L. Martini[1], S. Pezzini[1,3], V. Miseikis[1,3], U. Starke[4], F. Fabbri[1,3, ‡], C. Coletti[1,3,*]*

[1] Center for Nanotechnology Innovation @NEST, Istituto Italiano di Tecnologia, Piazza San Silvestro 12, I-56127 Pisa, Italy

[2] NEST, Scuola Normale Superiore, Piazza San Silvestro 12, I-56127 Pisa, Italy

[3] Graphene Labs, Istituto italiano di tecnologia, Via Morego 30, I-16163 Genova, Italy

[4] Max-Planck-Institut für Festkörperforschung, Heisenbergstraße 1, D-70569 Stuttgart, Germany

[‡] present address: NEST, Istituto Nanoscienze – CNR, Scuola Normale Superiore, Piazza San Silvestro 12, I-56127 Pisa, Italy

**\*E-mail:** camilla.coletti@iit.it

**Abstract**

The combination of the exciting properties of graphene with those of monolayer tungsten disulfide (WS$_2$) makes this heterostack of great interest for electronic, optoelectronic and spintronic applications. The scalable synthesis of graphene/WS$_2$ heterostructures on technologically attractive substrates like SiO$_2$ would greatly facilitate the implementation of novel two-dimensional (2D) devices. In this work, we report the direct growth of monolayer WS$_2$ via chemical vapor deposition (CVD) on single-crystal graphene arrays on SiO$_2$. Remarkably, spectroscopic and microscopic characterization reveals that WS$_2$ grows only on top of the graphene crystals so that the vertical heterostack is selectively obtained in a bottom-up fashion. Spectroscopic characterization indicates that, after WS$_2$ synthesis, graphene undergoes compressive strain and hole doping. Tailored experiments show that such hole doping is caused by the modification of the SiO$_2$ stoichiometry at the graphene/SiO$_2$ interface during the WS$_2$ growth. Electrical transport measurements reveal that the heterostructure behaves like an electron-blocking layer at large positive gate voltage, which makes it a suitable candidate for the development of unipolar optoelectronic components.

## 1. Introduction

In the last few years, van der Waals heterostructures (vdWHs) based on graphene and transition metal dichalcogenides (TMDs) have emerged as promising candidates for a wide number of applications. TMDs have unique properties in the 2D limit, such as indirect-to-direct band gap transition [1], broad and strong absorption in the spectral range from ultra-violet to visible, large exciton binding energy [2], well-defined valley degrees of freedom and sizeable spin splitting of the valence band maximum (VBM) [3]. In particular, monolayer TMD/graphene stacks are vdWHs of interest since they combine the high carrier mobility of graphene [4] as well as the strong light-matter interactions of single layer TMD [5]. Indeed, such heterojunctions have already been exploited in functional architectures, such as photodetectors [6–14] and optospintronic devices [15,16]. Furthermore, monolayer TMDs can be used on top of graphene not only as active materials to create vdWHs with enhanced electrooptical properties, but also as passive encapsulants, to preserve graphene mobility [17]. However, to move towards realistic applications, it is fundamental to develop an entirely scalable approach for the fabrication of vdWHs. At present, chemical vapor deposition (CVD) is the most suitable technique for the scalable synthesis of highly-crystalline 2D heterostructures [18,19]. The direct synthesis of $WS_2$ on graphene reduces the number of transfer steps, simplifying the fabrication process and is an ideal approach to obtain an atomically sharp interface [20].

To date, few works have demonstrated the direct synthesis of $WS_2$ on polycrystalline CVD graphene to obtain vertical [21] or lateral [10,13] heterostructures, the latter for the fabrication of photodetectors. Rossi et al. have shown the patterned synthesis of $WS_2$ on epitaxial graphene on silicon carbide (SiC) and the photodetection performance of such heterostructure [14]. However, no work has yet reported the direct synthesis of $WS_2$ on scalable high-mobility single-crystal CVD graphene on $SiO_2$, an appealing platform for the development of optoelectronic devices. Also, the effect of $WS_2$ growth on the properties of the underlying graphene crystal has been to date overlooked. Indeed, a thorough understanding of the influence of the direct CVD growth of $WS_2$ on the electronic and structural properties of single-crystal graphene is instrumental for identifying optimal paths to obtain performing and atomically sharp vdWHs.

In this work, we demonstrate the direct CVD growth of a single WS$_2$ layer on graphene single crystal arrays, deterministically grown via CVD on copper (Cu) foil and transferred on a technologically relevant substrate (i.e., SiO$_2$) [18,22]. The scalable synthetic approach demonstrated here is suitable for the implementation of microelectronic and optoelectronic devices. We thoroughly characterize the heterostack via Raman spectroscopy, photoluminescence (PL) spectroscopy, scanning electron microscopy (SEM), X-ray photoemission spectroscopy (XPS) and electrical transport measurements. Raman spectroscopy indicates that, upon WS$_2$ growth, graphene exhibits compressive strain and p-type doping, the latter caused by decomposition of the SiO$_2$ substrate as confirmed by XPS measurements. Furthermore, we present electrical transport measurements performed on the synthesized heterostack. The resistance as a function of gate voltage shows a significant electron-hole asymmetry, due to the presence of sulfur vacancies in WS$_2$ which form a trap level for graphene electrons. This hole-transporting/electron-blocking property could be conveniently employed for the development of unipolar optoelectronic components.

## 2. Methods

Graphene single crystal arrays [23–25] with a lateral size of about 200 µm were deterministically synthesized via CVD on electropolished copper (Cu) foils (Alfa Aesar, 99.8%) following the procedure described by Miseikis et al. [18] The Cu foils were selectively patterned using chromium (Cr) disks, which act as nucleation seeds for graphene crystals. Graphene was then synthesized at a temperature of 1060 °C inside a cold-wall CVD system (Aixtron BM) under methane, hydrogen and argon flow. The crystals were subsequently transferred on SiO$_2$/Si substrates (285 nm thick SiO$_2$ layer on p-doped Si, Sil'tronix) using a semi-dry procedure [18,22]. Specifically, a poly(methyl methacrylate) (PMMA AR-P 679.02 Allresist GmbH) film was used to support the graphene single crystals while detaching them from Cu using electrochemical delamination. The PMMA-coated graphene array was subsequently aligned with the target Si/SiO$_2$ substrate using a micromechanical stage and finally deposited on it. More details about the graphene growth and the transfer technique can be found in the Supplementary Information.

WS$_2$ was grown directly on graphene on SiO$_2$ via CVD from solid precursors, i.e., tungsten trioxide WO$_3$ and sulfur S. The process was performed in a 2.5" horizontal hot-wall furnace. As sketched in Figure S2, the furnace comprises a central hot-zone, where a crucible with the WO$_3$ powder was placed 20 mm away from the growth substrate, and an inlet zone heated by a resistive belt, in which the S powder was positioned, in order to separately control its temperature. S was evaporated and then carried by an argon flux to the centre of the furnace where it reacted with $WO_3$ directly on the sample surface at a temperature of 900 °C and at a pressure of ~5 x $10^{-2}$ mbar, see Supplementary Information for additional details. The growth time necessary for a full coverage of the graphene crystals was 20 minutes. Partial growths, instrumental to assess the size and orientation of the $WS_2$ crystals, were carried out decreasing the growth time down to 5 minutes. Such partial growths yielded a $WS_2$ coverage of graphene of about 80% (see Figure 1(b)).

The exfoliated h-BN flakes used in this work were purchased from HQ Graphene.

Thermal annealing experiments were performed in an ultra-high vacuum (UHV) chamber (base pressure of 2 x $10^{-10}$ mbar). Temperature ramp up/down rates of 1°C/min were used and the target temperatures were maintained for 10 hours.

Raman spectroscopy was used to assess crystal quality, doping and strain of graphene and was performed together with PL to characterize quality and thickness of $WS_2$. Both Raman and PL measurements were carried out with a Renishaw InVia spectrometer equipped with a 532 nm laser with a spot size of ~1 μm. The power used was 1 mW.

Scanning electron microscopy (SEM) was performed to study the morphology of each sample. A Zeiss Merlin microscope and electrons with an accelerating voltage of 5 kV were used.

X-Ray photoelectron spectroscopy (XPS) measurements were performed at room temperature with a Mg Kα anode coupled to a Phoibos150 electron analyzer from SPECS GmbH. The photoemission angle used was 60° with respect to the surface normal in order to increase the surface sensitivity of the measurements. The binding energies of the peaks reported in the text were referenced to the energy of graphene $sp^2$ carbon set at 284.5 eV.

To investigate the transport characteristics of graphene after $WS_2$ growth, multi-terminal field-effect transistor (FET) devices were fabricated on $WS_2$/graphene/$SiO_2$. Electric-field effect measurements were performed at room temperature using a Keithley 2450 sourcemeter with a micro-probe station in air. By applying a constant current of 1 μA between the external electrodes, we measured the voltage drop along the device as function of the applied back-gate voltage ($V_{bg}$).

## 3. Results and discussion

*3.1 $WS_2$/graphene scalable heterostructures*

Figure 1(a) shows the typical graphene single-crystal array grown by deterministic seeding and transferred on SiO$_2$ [18] adopted to perform CVD growth of WS$_2$. This synthetic approach of graphene has significant prospects for scalability and is flexible, i.e. single crystals of different size and spacing can be obtained. While reducing transfer related problems (i.e., tears and breaks are reduced when transferring smaller tiles rather than continuous wafers), this approach also allows to have high-mobility graphene exactly where needed [18]. Indeed, the positioning of the crystal can be flexibly designed accordingly to the mask of any final optoelectronic device.

In initial experiments, partial growth of WS$_2$ was performed to assess dimension and orientation of the crystals of the synthesized TMD on graphene. A SEM image of a portion of a graphene crystal covered with WS$_2$ flakes is displayed in Figure 1(b). The WS$_2$ triangular crystals have dimensions of a few hundred nanometers and most of them are merged with adjacent ones. A higher density of WS$_2$ flakes is clearly visible along the graphene wrinkles, indicating that the morphology of the underlying graphene layer, in particular the presence of defects, strongly affects WS$_2$ nucleation. WS$_2$ crystals present only two different orientations related to each other by a rotation of 60°. This suggests the existence of an epitaxial relation between WS$_2$ and graphene, which was already reported for monolayer WS$_2$ directly grown on 2D materials [19,21,26,27].

A typical Raman spectrum of the synthetized WS$_2$, probed with a 532 nm laser, is displayed in Figure 1(c). The Raman feature around 350 cm$^{-1}$ is the convolution of the four peaks 2LA($M$)-E$^2_{2g}$($\Gamma$), E$^1_{2g}$($M$), 2LA($M$), and E$^1_{2g}$($\Gamma$), while the peak A$_{1g}$($\Gamma$) at 417.1 cm$^{-1}$ is a first-order mode corresponding to out-of-plane oscillations of atoms. When using a specific excitation wavelength, the second-order phonon mode 2LA($M$) is more prominent than the in-plane phonon mode E$^1_{2g}$($\Gamma$) and the 2LA($M$)/A$_{1g}$($\Gamma$) intensity ratio is used as indicative parameter of the thickness of WS$_2$ [28]. A 2LA($M$)/A$_{1g}$($\Gamma$) intensity ratio above 2, as in this case, is typical of monolayer WS$_2$ crystals [28]. Photoluminescence spectra reported in Supplementary Information Figure S3 further confirm that WS$_2$ is single-layer.

Remarkably, panel (b) suggests that WS$_2$ crystals only grow on the graphene flakes and not on the SiO$_2$ substrate, probably owing to the desorption of oxygen from SiO$_2$ which hinders the sulfurization of the WO$_3$ precursor (for further information see Section 3.4 and Figure S8 in Supplementary Information). In order to confirm what is observed by SEM, we performed scanning Raman spectroscopy experiments. Figure 1(d) shows a Raman map of the intensity of the 2LA($M$) mode over 2x2 graphene single-crystal array after WS$_2$ synthesis. Indeed, WS$_2$ growth seems to be strongly favored on the graphene substrate: no WS$_2$ Raman signatures are detected on SiO$_2$ (for additional Raman analysis see Supplementary Information Figure S3). The selective growth of WS$_2$ crystals on the graphene flakes represents a clear advantage for the fabrication of devices as the entire vertical heterostructure is deterministically obtained in arrays without the need of any top-down post-processing.

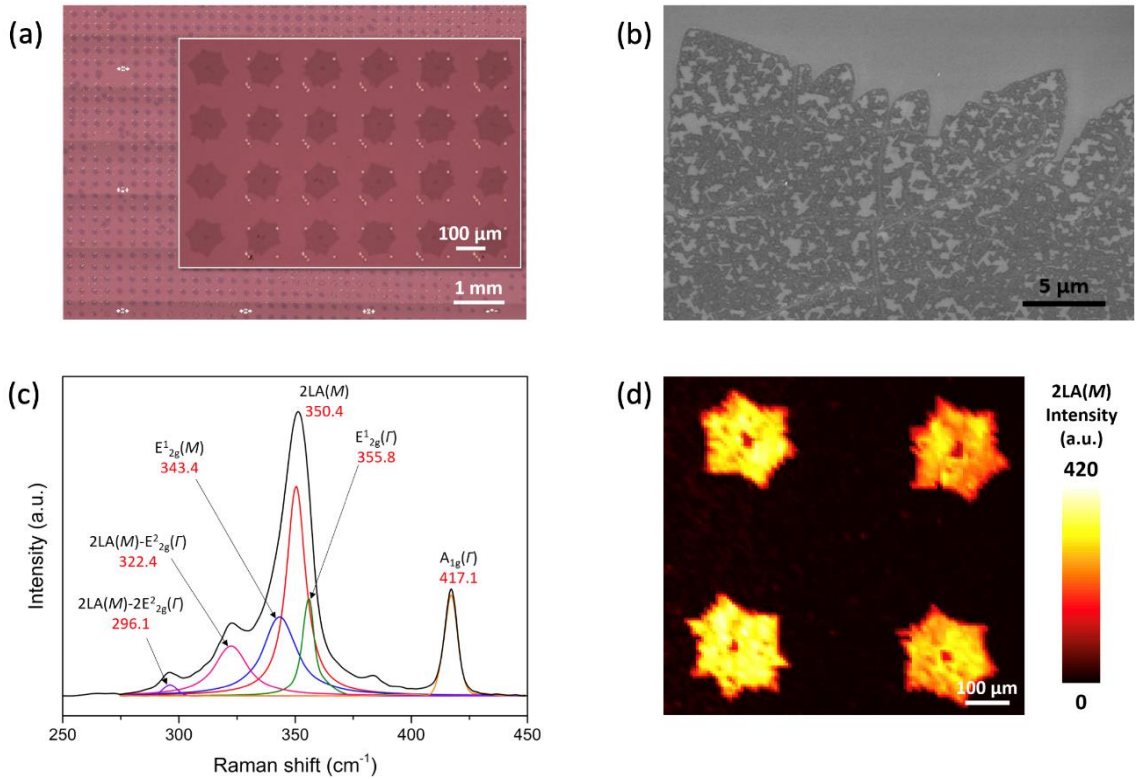

**Figure 1.** (a) Optical image of a graphene single-crystal array grown by deterministic seeding and transferred on SiO$_2$. (b) SEM image of a portion of a graphene crystal transferred on SiO$_2$ and covered with WS$_2$ flakes. (c) Raman spectrum of WS$_2$ on CVD graphene. (d) Raman map of the intensity of the 2LA(*M*) mode over 2x2 graphene single-crystal array after WS$_2$ synthesis.

*3.2 Effect of the direct synthesis of WS$_2$ on graphene: doping and strain*

In order to investigate the full applicative potential of the heterostack and to further develop the materials synthesis, it is crucial to characterize the properties of the CVD graphene crystals after WS$_2$ growth, something that to date has not yet been exhaustively addressed. We adopt Raman spectroscopy as a powerful tool yielding insights on the structural and electronic properties of graphene. In the case of single-layer graphene, any shift of the two prominent G and 2D Raman peaks is attributable to strain [29,30] and/or doping [31–34] in the material. In

Figure 2(a) we show the Raman spectra of graphene before and after $WS_2$ synthesis. A considerable stiffening of both characteristic modes, G (26.5 cm$^{-1}$ blue-shift) and 2D (35 cm$^{-1}$ blue-shift), is consistently observed after $WS_2$ growth. Notably, no increase in the intensity of the negligible D-peak is observed, indicating that the growth process does not induce significant defects in the crystal structure of the underlying graphene. The significant blue-shift of the 2D peak observed after $WS_2$ growth typically indicates that the majority carriers in graphene are holes [33]. A correlation plot of the G and 2D peak positions ($\omega_G$ and $\omega_{2D}$ respectively) – instrumental to disentangle the strain and doping contribution [35,36] – is shown in panel (b). The two solid lines are the directions along which the strain-induced (purple line) and the hole doping-induced (orange line) shifts are expected [35]. The intersection of the two lines ($\omega^0_G$, $\omega^0_{2D}$) = (1583, 2678) cm$^{-1}$ represents the case of neutral and unstrained graphene [33]. The data plotted in panel (b) are extracted from representative 15x15 µm$^2$ analyzed areas. Remarkably, the graphene crystals transferred by semi-dry process (black data points) are in close proximity to the unstrained and undoped reference, differently from wet-transferred CVD graphene where residual doping >10$^{12}$ cm$^{-2}$ – caused by the transfer process – is typically found [37–40]. After $WS_2$ growth, the graphene crystals present compressive strain and hole doping estimated to be about 0.6 % and 10$^{13}$ cm$^{-2}$, respectively (red data points). The increase in doping is also confirmed by the strong reduction of the intensity of the 2D peak with respect to the intensity of the G peak (Figure 2(c)) [31,41,42]. Furthermore, the significant enlargement of the 2D full-width-at-half-maximum (FWHM) value upon $WS_2$ growth (Figure 2(d)), from a remarkably low 24 cm$^{-1}$ to about 37 cm$^{-1}$, suggests an increase of the strain fluctuation in the graphene crystal [43], a possible limiting factor for the carrier mobility in this vdWH.

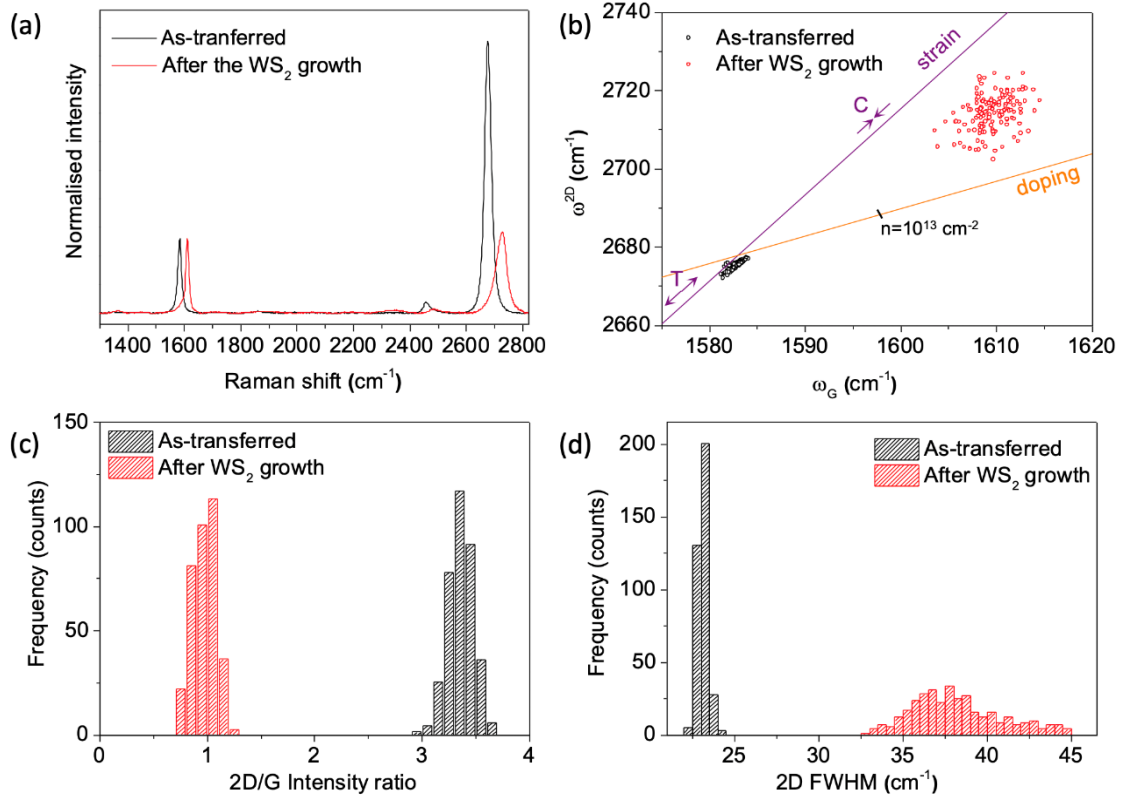

**Figure 2.** (a) Raman spectra of graphene before (black) and after (red) WS$_2$ growth. (b) Correlation plot of the G vs. 2D Raman modes frequencies recorded on graphene before and after WS$_2$ growth. The purple line indicates the $\omega_G$-$\omega_{2D}$ for charge-neutral graphene under compressive (C) or tensile (T) strain, as predicted in ref.[35]. The orange line represents the $\omega_G$-$\omega_{2D}$ correlation for p-type doping. The purple dotted line, which indicates the $\omega_G$-$\omega_{2D}$ for a doping of about $10^{13}$ cm$^{-2}$, is inserted as a guide for the reader's eyes. (c) Histogram of the intensity ratio between the 2D peak and the G peak before (black) and after (red) WS$_2$ growth. (d) Histogram of the 2D peak FWHM before (black) and after (red) WS$_2$ growth. Both the histograms refer to a 15x15 µm$^2$ representative area.

*3.3 Effect of thermal annealing on CVD single-crystal graphene properties*

In order to assess the influence of the high-temperature annealing (carried out during the growth process) on graphene strain and doping, annealing processes were performed in a controlled atmosphere. One sample was annealed at 900°C (i.e., WS$_2$ growth temperature) under Ar flux, in order to replicate the conditions experienced by the sample during the CVD growth of WS$_2$ (same time and pressure, but without the presence of solid precursors). Two other samples were annealed in UHV, one at 300°C and the other at 500°C.

We find that when annealing the sample at 300 °C, the Raman fingerprint of graphene is only slightly affected while more substantial changes are observed for the 500 °C and 900 °C annealing stages (see Figure 3(a) and Supplementary Information Figure S5). Indeed, the $\omega_{2D}$ - $\omega_G$ correlation plot indicates a little increase of hole doping for the 300 °C annealing and the emergence of a stronger doping entirely comparable to that observed after WS$_2$ synthesis (i.e., 10$^{13}$ cm$^{-2}$) for higher annealing temperatures. Remarkably, the central values of the dataset distributions relating to as-transferred and annealed graphene, have almost the same projection onto the strain axis. This indicates that thermal treatments carried out in these specific experimental conditions contribute mostly to the doping, which reaches its maximum value already at 500°C. This finding is qualitatively in agreement to what was reported for polycrystalline CVD graphene heated under He atmosphere by Costa et al. [44], although in the correlation plot we retrieve smaller data spreads for the pristine and low-temperature annealed samples, as a consequence of the higher quality and homogeneity of single-crystal graphene.

These data also suggest that the compressive strain observed in graphene after WS$_2$ growth is largely due to the interaction with the overlying material. The model we propose is that, in virtue of its negative thermal expansion coefficient [45–47], graphene lattice parameters get reduced during the heating. When WS$_2$ nucleates on such a compressed lattice, it may find a more stable configuration for forming a coincidence lattice with graphene, locking graphene in the strained configuration. In a previous work we found that WS$_2$ on epitaxial graphene, which is characterized by a compressed lattice, form a (7×7) on a (9×9) superperiodicity with graphene [48]. Admitting that as a sort of natural coincidence lattice, graphene should shrink by about 0.45%, a value in line with what is extracted from the Raman analysis above.

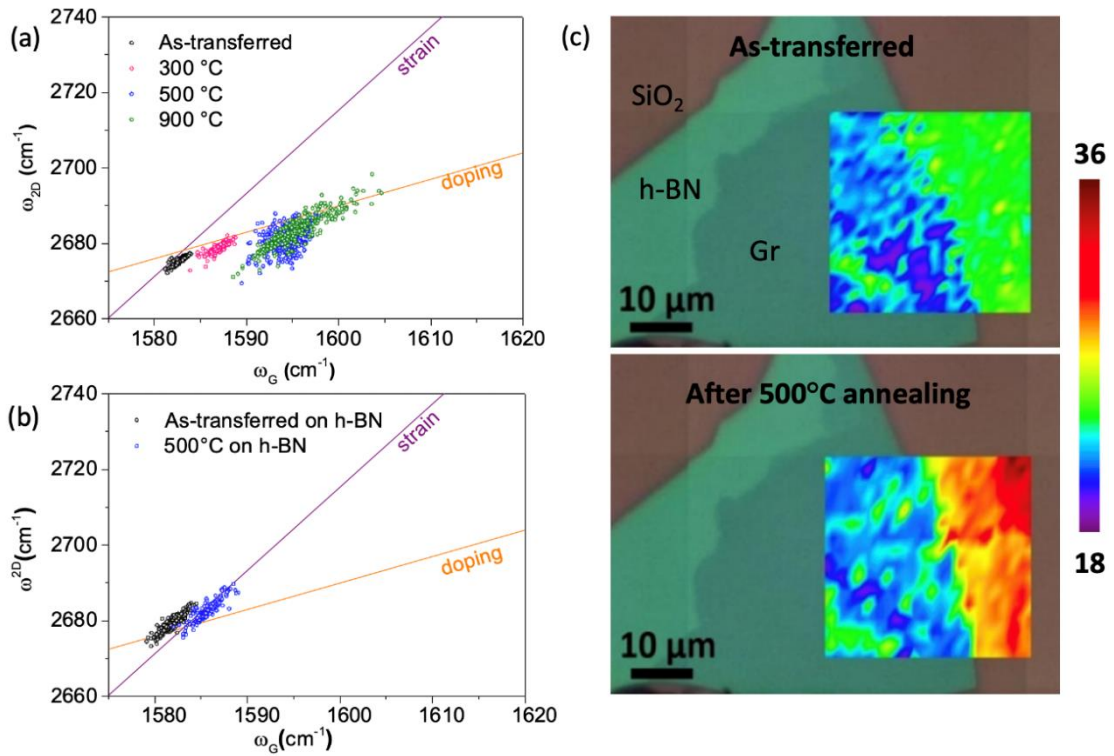

**Figure 3.** (a) Correlation plot of the $\omega_G$-$\omega_{2D}$ Raman modes measured on pristine graphene and after each annealing (300°C in pink, 500°C in blue and 900°C in green). The purple dashed line indicates a doping of about $10^{13}$ cm$^{-2}$. (b) Correlation plot of the $\omega_G$-$\omega_{2D}$ Raman modes frequencies recorded on graphene on h-BN before (black) and after (blue) annealing at 500°C. (c) Raman maps of 2D peak FWHM (cm$^{-1}$) before (top) and after (bottom) annealing superimposed to the optical image (the green flake is exfoliated h-BN).

*3.4 The origin of doping in annealed CVD graphene on SiO$_2$*

Having clarified that graphene hole doping upon WS$_2$ growth is a result of the thermal annealing of the sample, we now focus on identifying the physico-chemical origin of such doping. To date, a number of works have attributed hole doping of annealed exfoliated graphene on SiO$_2$ to ambient H$_2$O and O$_2$ molecules adsorbed owing to thermal-induced structural deformation or defects [44,49–52]. Other works have instead suggested that the hole doping is mostly induced by the SiO$_2$ substrate, that upon annealing experiences an enhanced coupling with graphene [44,53,54]. In order to understand whether the doping is substrate- or atmosphere-induced, we carried out a comparative experiment. A large CVD graphene crystal was transferred partly on an exfoliated hexagonal boron nitride (h-BN) flake (∼ 20 nm thick) and partly directly on top of SiO$_2$ and then annealed at 500°C in UHV. h-BN, indeed, is well-known as an ideal and effective encapsulant for graphene [55], capable of screening doping from the substrate and atmosphere when used as a bottom- or top- encapsulant, respectively [56].

Raman characterization, performed after extracting the sample from the vacuum chamber, showed that the annealing did not induce either significant doping or strain (see correlation plot in panel (b)) in the graphene portion placed on top of h-BN. In contrast, the portion on SiO$_2$ shows the same distribution as the blue one in panel (a). Furthermore, Raman maps in Figure 3(c) report a significant broadening of the 2D peak for annealed graphene on SiO$_2$, while no broadening is observed upon the annealing of graphene on h-BN. This result rules out molecular adsorption due to air exposure as a possible cause for the measured doping. If the doping was due to the environment, we would expect a similar hole doping level for both the graphene/h-BN and the graphene/SiO$_2$ region. Furthermore, differently from previous works [44,49,50-52] in our system WS$_2$ acts as a top-encapsulant, hindering atmospheric-induced doping of graphene. Therefore, the doping originates from graphene/SiO$_2$ interface states activated by thermal annealing. XPS measurements performed on a sample with CVD graphene on SiO$_2$ before and after annealing at 500°C in UHV corroborate this thesis.

We display the results of those measurements in Figure 4. Panels (a) and (b) show a clear decrease of the binding energy for the O 1s and Si 2p peaks upon annealing, which indicates an oxygen loss from the top layers of the substrate, yielding a graphene/SiO$_{(2-\delta)}$ interface. Such an under stochiometric interface generates a polarization field the net effect of which is the p-type doping of graphene. The apparent increase of the peak intensity after the annealing is attributed to carbon contaminants, which are desorbed from the surface. This is well visible when looking at the C 1s peak, reported in Figure 4(c). The C 1s peak shifts as a whole by about 450 meV, again towards lower binding energies. However, we identify at least four components in the peak before annealing and three after annealing. It is well known that after standard cleaning procedures [40,57–59] residues of the PMMA layer used during transfer are still present on the graphene surface. We therefore assign the three components of the C 1s peak considerably reduced upon thermal treatment (i.e., due to evaporation of the polymer from the surface) to polymeric residues. The component at lowest binding energy which is not affected by the annealing is assigned to sp$^2$ bound carbon, i.e. graphene. The sp$^2$ carbon component shows a net shift by about 140 meV towards lower binding energy. Since for small deviations the shift of the binding energy and the graphene doping are linearly related [59], we can conclude that the XPS data is qualitatively in agreement with the Raman data, ultimately confirming a displacement towards higher p-type doping of the annealed graphene layer (more details can be found in Supplementary Information Figure S6).

In virtue of what is shown in Figure 3, we propose inserting a protective layer (possibly another 2D material) between graphene and the substrate to avoid p-type doping of graphene during CVD growth or other high-temperature process.

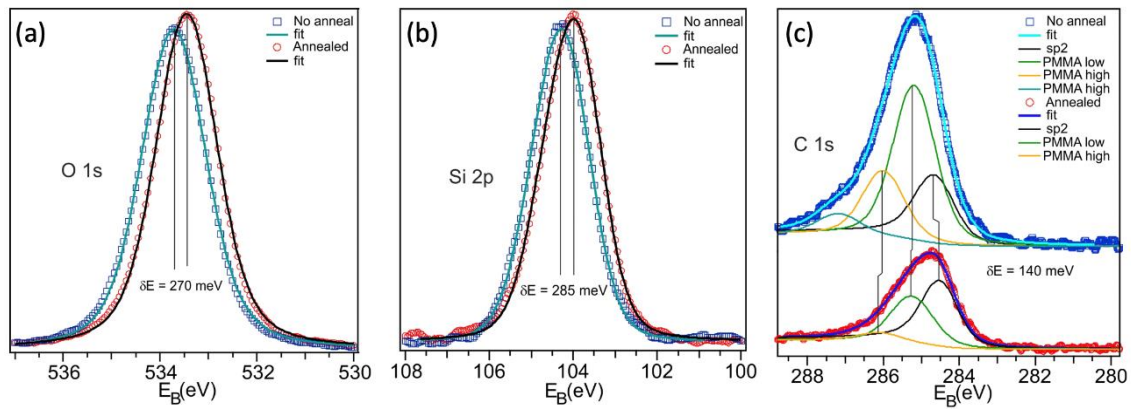

**Figure 4.** XPS measurements. O 1s (a), Si 2p (b) and C 1s (c) spectra of a sample with CVD graphene on SiO$_2$ before and after annealing at 500°C.

*3.4 Electrical transport measurements*

The electrical proprieties of the synthesized $WS_2$/graphene vdWH were assessed via four-contact electrical measurements (Figure 5(a)) in ambient conditions. A typical $WS_2$/graphene multi-terminal field-effect transistor (FET) device is shown in Figure 5(b) (details on the device fabrication can be found in Supplementary Information Figure S7). In Figure 5(c) we show the resistance of the vdWH (red curve) and that of a reference pristine CVD single-crystal graphene (black curve) as a function of the voltage ($V_{bg}$) applied to the back gate (p-doped Si wafer with 285 nm of thermal oxide). The two devices have equal geometry and have undergone the same fabrication steps, making their resistance curves directly comparable. To make the comparison between the two transfer characteristics easier, along the x-axis we put the $V_{bg}$ relative to the charge neutrality points (CNPs) of the two devices. The CNP in the after-$WS_2$-growth case, indeed, was measured at higher positive gate voltages (~20.5 V for the least doped measured sample) with respect to the graphene-only case (~12.4 V), confirming again that the CVD process for $WS_2$ synthesis dopes graphene with holes. In particular, the hole density at $V_{bg}$ = 0 for this particular device, estimated by the CNP shift, is $n_h$~1.5 x $10^{12}$ $cm^{-2}$.

While the black curve shows the typical symmetric peak of graphene, a marked electron-hole asymmetry is observed for the vdWH, with the red trace almost pinned to the resistance value at the CNP. This is assigned to the presence of sulfur vacancies in $WS_2$, identified via XPS analysis (see Supplementary Information Figure S8). These vacancies in the $WS_2$ layer trap the electrons induced by gating in the contiguous graphene layer (Figure 5(d)). The saturation of the resistance curve for the vdWH can be easily attributed to having a small density of electrons in the graphene sheet and contributing to the conductance, despite the large gate voltage applied, with the 'missing' carriers occupying $WS_2$ defect states. A similar (although less marked) effect of hole-electron asymmetry in $WS_2$/graphene vdWH was reported by Avsar et al. [15] using mechanically exfoliated $WS_2$, which are free from grain boundaries and with a low edges/area ratio. Since flake's edges are the points where sulfur vacancies are mostly concentrated in TMDs [60], the enhancement of the electron-hole asymmetry in our experiment can be easily explained. This asymmetry in the conduction of holes and electrons can be conveniently exploited in a number of applications ranging from energy conversion to optical detection. For example, in solar cells, this heterostack used as an anode contact could help to optimize the collection of the photogenerated holes, blocking the electrons at the same time [61,62]. Furthermore, considering the extremely fast charge transfer already demonstrated between the two materials

[63,64] this deterministically grown entirely scalable vdWH could be exploited to develop unipolar devices for optoelectronics and optospintronics.

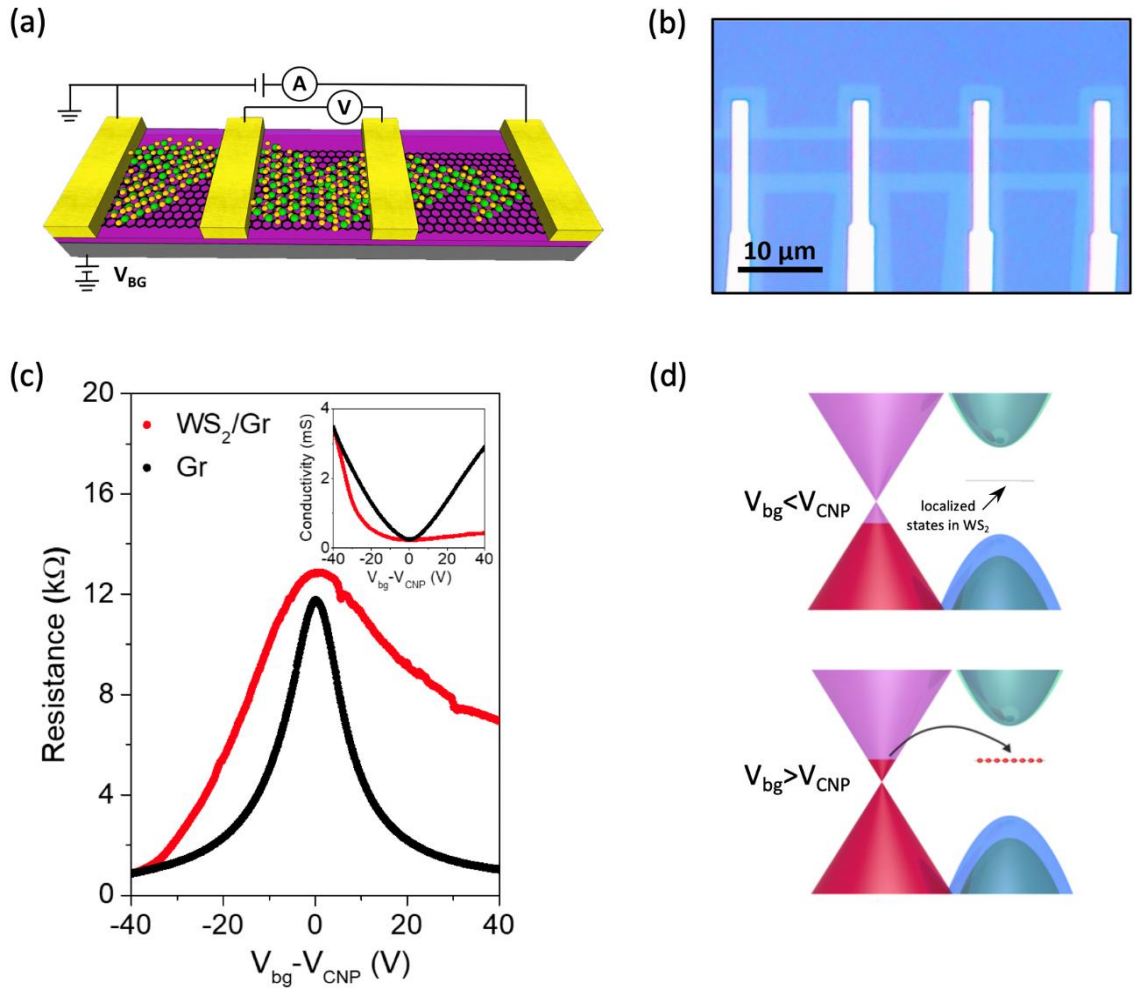

**Figure 5.** (a) Sketch of the measurement geometry. (b) Optical image of a device fabricated on the WS$_2$/graphene heterostructure. (c) WS$_2$/graphene/SiO$_2$ (red curve) and graphene/SiO$_2$ (black curve) resistance as a function of the back-gate voltage relative to the CNPs of the two devices. Inset: WS$_2$/graphene/SiO$_2$ (red curve) and graphene/SiO$_2$ (black curve) conductivity as a function of the back-gate voltage relative to the CNPs of the two devices. (d) Band schematic of WS$_2$ and graphene when the Fermi level is moved by the back gate below (top) and above (bottom) the CNP of graphene.

## 4. Conclusions

In this work, we demonstrate the direct synthesis of monolayer WS$_2$ on single-crystal CVD graphene arrays. We observe a selective growth of WS$_2$ crystals on graphene, a clear advantage for the fabrication of devices as the vertical heterostructure is deterministically obtained in arrays without the need of any top-down post-processing. The presented heterostructure has the advantages of being fully scalable and compatible with a silicon technology platform. We also investigate the structural and electrical properties of graphene after WS$_2$ growth. The graphene substrate turns out to be affected by hole doping and compressive strain. While the strain is attributed to the interaction with the WS$_2$ overlayer, thermal treatment experiments allow us to assign the doping of graphene to SiO$_2$ interface states activated by the high temperatures during the TMD synthesis. We demonstrate that a protective layer such as h-BN placed between graphene and the SiO$_2$ substrate before growth is instrumental to avoid the doping. Finally, we show that the heterostructure behaves like an electron-blocking layer, which might be suitable for the development of unipolar components in optoelectronics.

**Acknowledgements**

The authors would like to thank Mauro Gemmi, Simona Pace and Antonio Rossi from CNI@NEST for support in TEM imaging.

The research leading to these results has received funding from the European Union's Horizon 2020 research and innovation program under grant agreement No. 785219 – GrapheneCore2.

# Supplementary Information

## Deterministic direct growth of WS$_2$ on CVD graphene arrays

*G. Piccinini[1,2], S. Forti[1], L. Martini[1], S. Pezzini[1,3], V. Miseikis[1,3], U. Starke[4], F. Fabbri[1,3, ‡], C. Coletti[1,3, *]*

[1] Center for Nanotechnology Innovation @NEST, Istituto Italiano di Tecnologia, Piazza San Silvestro 12, I-56127 Pisa

[2] NEST, Scuola Normale Superiore, Piazza San Silvestro 12, I-56127 Pisa, Italy

[3] Graphene Labs, Istituto italiano di tecnologia, Via Morego 30, I-16163 Genova

[4] Max-Planck-Institut für Festkörperforschung, Heisenbergstraße 1, D-70569 Stuttgart, Germany

[‡] present address: NEST, Istituto Nanoscienze – CNR, Scuola Normale Superiore, Piazza San Silvestro 12, I-56127 Pisa, Italy

**\*E-mail:** camilla.coletti@iit.it

**Graphene growth and transfer**

Graphene single crystals arrays were synthesized via chemical vapor deposition (CVD) at a pressure of 25 mbar inside a 4-inch cold-wall CVD system (Aixtron BM) [1]. Electropolished copper (Cu) foils (Alfa Aesar, 99.8%) were used as catalytic substrates. Nucleation sites, with a diameter of 5 µm, were patterned on the foils by optical lithography and thermal evaporation of 25 nm of chromium, followed by lift-off. The substrates were then annealed in argon atmosphere for 10 minutes within the CVD system to preserve surface oxidation. Graphene was subsequently synthesized at a temperature of 1060 °C flowing over the sample methane, hydrogen and argon at 1 sccm, 100 sccm and 900 sccm, respectively [2]. The growth time was 20 minutes. After that the chamber was cooled down in argon/hydrogen atmosphere to a temperature of 120 °C before removing the sample from the reactor. The graphene crystals were then transferred on Si substrates with a 285-nm-thick SiO$_2$ layer by means of a semi-dry procedure consisting in reinforcing the graphene layer with a poly(methyl methacrylate) (PMMA AR-P 679.02 Allresist GmbH) film and in detaching it from copper using electrochemical delamination[1] (Figure S1). We first attached a polydimethylsiloxane (PDMS) frame to the copper/graphene/PMMA stack in order to handle the polymeric foil once released from the copper. After delamination, the graphene/PMMA was rinsed in de-ionized (DI) water and deposited on the substrate using a micromechanical stage. During the transfer, the substrate was heated at 120 °C to increase the graphene-substrate adhesion. The PMMA was finally removed in acetone (ACE) and isopropanol (IPA). For improved cleaning of the sample from polymer residues, PMMA remover (AR-P 600-71 Allresist GmbH) was also employed.

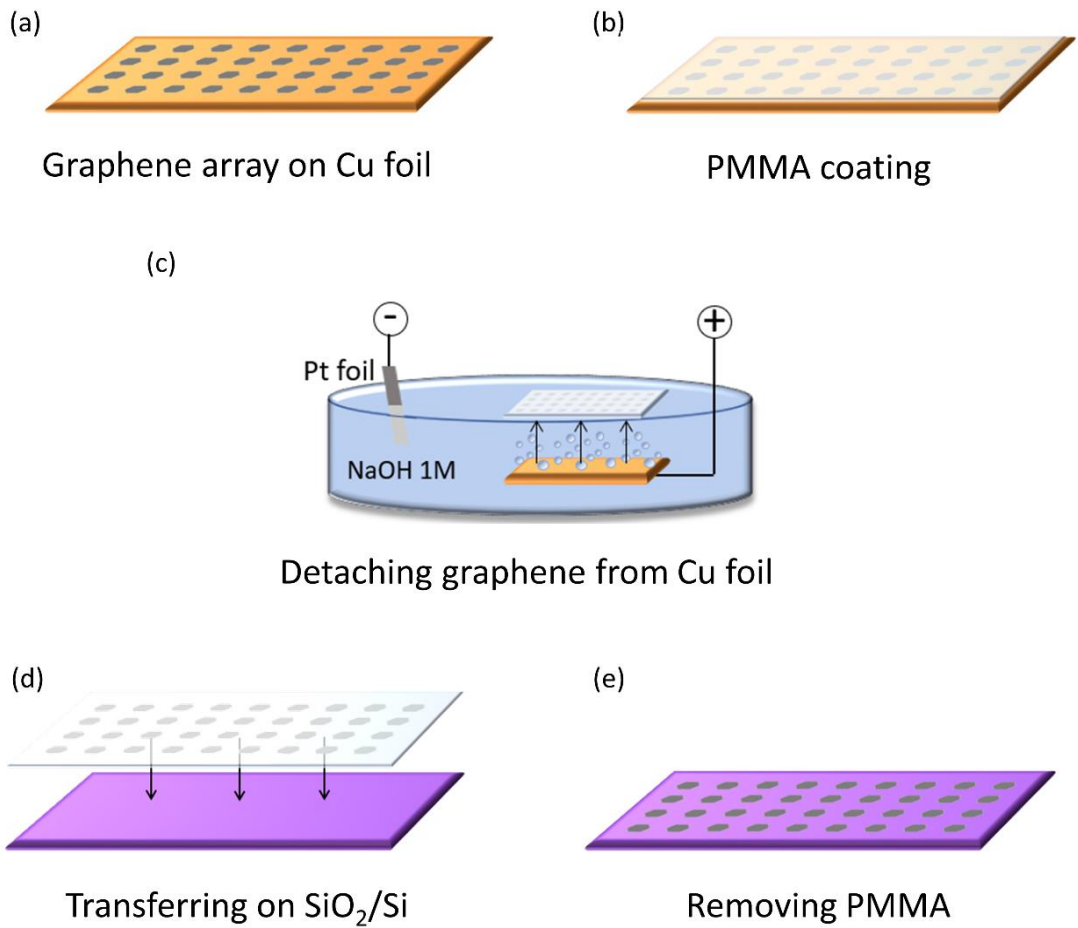

**Figure S1.** Sketch of the semi-dry transfer procedure for CVD graphene single-crystal arrays.

**WS$_2$ growth**

To synthesize WS$_2$ directly on the CVD graphene/SiO$_2$ substrate, a CVD approach was adopted, using as precursors WO$_3$ (Sigma Aldrich, 99.995%) and S (Sigma Aldrich, 99.998%) powders in a

1:50 ratio (3 mg of WO$_3$ and 150 mg of S). After the chamber was pumped down to a pressure of ∼ 5 x 10$^{-2}$ mbar, the temperature ramp-up was started with a rate of 5°C/min, paying attention to increase the pressure inside the chamber to a value high enough (4.6 mbar) to keep the sulfur solid. To do that, a flux of 500 sccm of Ar was flown during the temperature ramp-up. During the process, the temperature within the reaction zone was set to 900 °C, while the belt temperature was set to 200°C to evaporate sulfur. The heating belt was switched on only after reaching the growth temperature in the main zone of the reactor. At this point, the Ar flux was suddenly reduced to 8 sccm, which led the furnace pressure to drop immediately to 0.6 mbar. In these conditions, sulfur starts suddenly to evaporate and to sulfurize the WO$_3$ solid precursor. Finally, the furnace was naturally cooled down to room temperature and the sample was removed from the tube. It is worth noting that the coverage of the graphene flake strictly depends on the growth time. In order to achieve a complete coverage, the growth time was set to 20 minutes. In case of a partial growth, where the isolated WS$_2$ triangular monolayers are visible, the growth time needs to be reduced down to 5 minutes.

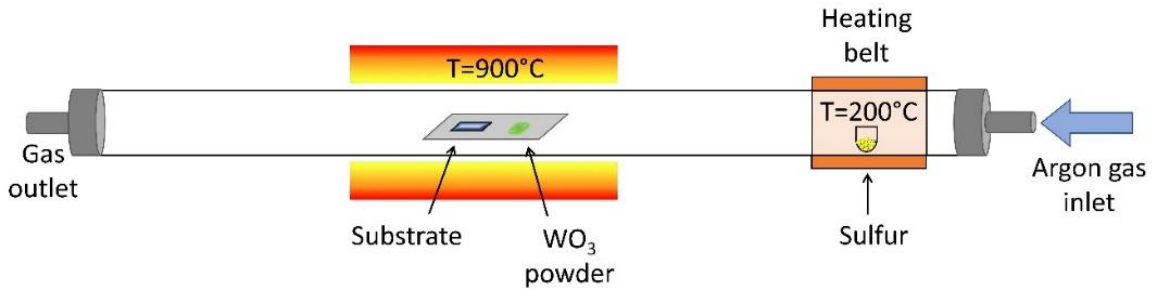

**Figure S2.** Sketch of the furnace for the CVD growth of WS$_2$.

**WS$_2$ photoluminescence and Raman spectroscopy**

A strong evidence of the monolayer nature of the synthesized WS$_2$ comes from photoluminescence (PL) spectra. The band gap structure of TMDs is highly dependent on their thickness [3]. Since monolayer WS$_2$ is a direct band gap semiconductor, only strong excitonic direct-transition (DT) emission located at ∼ 627 nm (1.98 eV) can be observed in its PL spectrum. As the WS$_2$ layer number increases, indirect-transition (IT) emissions should show up in the higher wavelength side of the DT emission [4]. In Figure S3(a) the PL spectrum is displayed. An intense PL response is visible and there is no sign of additional peaks relative to transitions at lower energy. Furthermore, the high-resolution photoluminescence mapping measurement in Figure S4(b) shows that graphene is homogenously covered with monolayer WS$_2$. The selective growth of WS$_2$ on graphene is confirmed by the Raman map of the intensity of the A$_{1g}$($\Gamma$) mode over a graphene single-crystal after WS$_2$ synthesis (Figure S3(c)). The total absence of this characteristic peak of $WS_2$ on $SiO_2$ suggests the possibility of defining the shape and position of the heterostructure by patterning graphene before $WS_2$ synthesis.

PL and Raman measurements were carried on with a Renishaw InVia system equipped with a 532 nm green laser and 100× objective lens, providing a spot size of ∼1 μm. The power used was 1 mW.

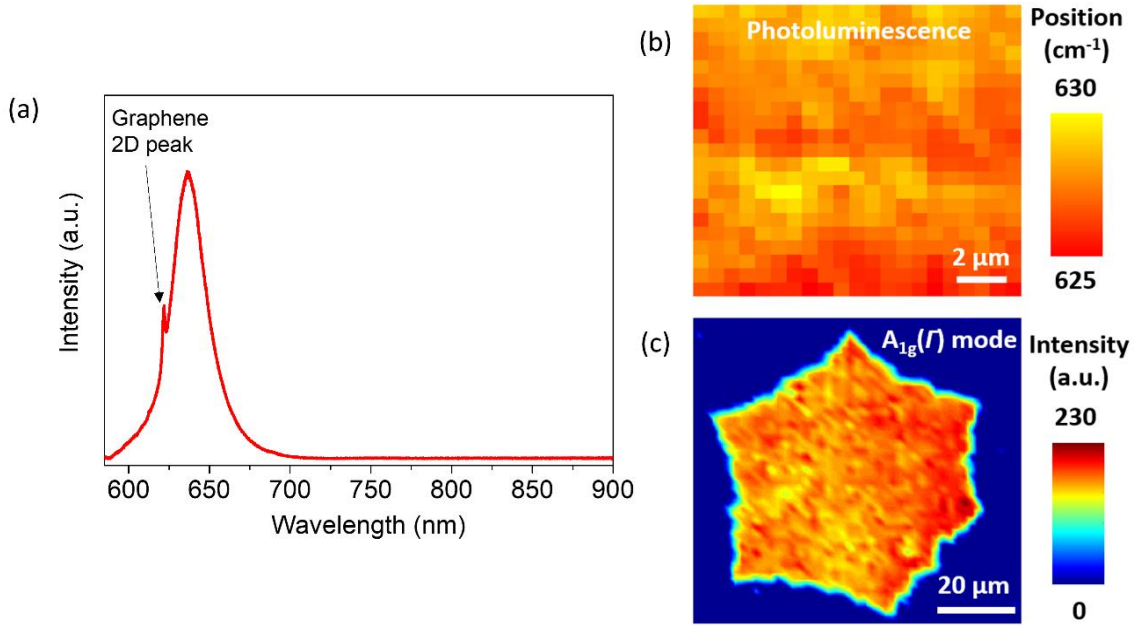

**Figure S3.** (a) Photoluminescence spectrum of $WS_2$. (b) High-resolution photoluminescence map of a 15x15 μm² area of the $WS_2$/graphene heterostructure, plotted by extracting the peak position. (c) Raman map of a graphene single-crystal after $WS_2$ synthesis, plotted by extracting the intensity of the $A_{1g}(\Gamma)$ mode.

**SEM and TEM measurements**

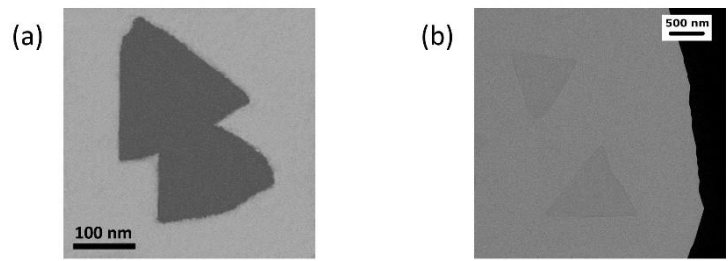

**Figure S4.** (a) High magnification SEM image of a homogeneous $WS_2$ monolayer, (b) TEM bright field image of $WS_2$ monolayers.

Figure S4(a) shows an high magnification image of $WS_2$ monolayer synthesized on CVD graphene, with a grow time of 5 minutes. The lateral size of the $WS_2$ crystals is about 200 nm in agreement with the low magnification image presented in the manuscript (Figure 1(b)).

In Figure S4(b) is displayed a bright field TEM analysis of $WS_2$ monolayers grown on CVD graphene. It is worth noting that for performing this analysis we carried out a dedicated 5 minutes long growth process on graphene transferred directly on gold TEM grids. This was necessary because the transfer process of the whole vdW heterostructure was always affected by polymeric residuals that did not allow a clear TEM imaging of the $WS_2$ flakes. Transmission electron microscopy was carried out on a Zeiss Libra 120 transmission electron microscope operating at 120 kV and equipped with an in-column Omega filter for energy filtered imaging.

**Raman characterization after different graphene annealing procedures**

Figure S5(a) shows how the graphene Raman spectrum changes after annealing at 300 °C and 500 °C in UHV and at 900 °C under Ar flux (this latter replicates the conditions experienced by the sample during the CVD growth of $WS_2$). The effect of doping is clear from both the blueshift [5–8] of the G peak and the intensity reduction [9] of the 2D peak, in particular when annealing above 300°C. A more quantitative analysis is given by the histogram in Figure S5(b). Furthermore, the significant broadening of the 2D peak upon annealing is the fingerprint of an increase of the strain fluctuation in the graphene crystal [10] (Figure S5(c)).

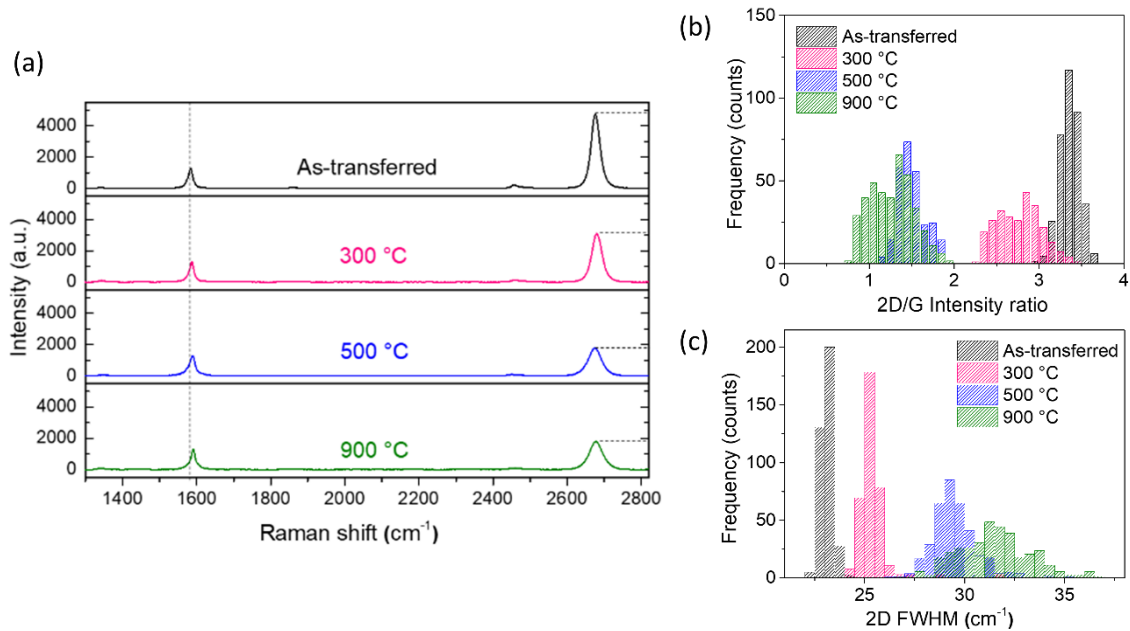

**Figure S5.** (a) Raman spectra of graphene before (black) and after annealing at 300°C (pink), 500°C (blue) and 900°C (green). (b) Histogram of the intensity ratio between the 2D peak and the G peak before and after each annealing. (c) Histogram of the 2D FWHM before and after each annealing. Both histograms refer to 15x15 µm² representative areas.

**Graphene on $SiO_2$ doping after annealing**

XPS measurements performed on a sample with CVD graphene on $SiO_2$ before and after annealing at 500°C in UHV suggest that the thermal treatment causes the loss of oxygen by the substrate at the interface with graphene. This is the reason for the high p-doping level observed in graphene on $SiO_2$ upon annealing. In Figure S6 the doping mechanism is displayed. The oxygen desorbed upon the high temperature treatment leaves an under stochiometric $SiO_{(2-\delta)}$ compound which exhibits an excess of negative charges, thereby polarizing the first layers of the interface. This causes the upshift of the graphene's π-bands and the consequent downward bending of the oxide HOMO (highest occupied molecular orbital) and LUMO (lowest occupied molecular orbital), owing to the polarization field induced by the oxygen desorption at high temperature. Once the equilibrium is achieved, the hole doping in graphene turns out to be increased according to the new position of the Fermi level, which remains pinned due to the high $SiO_2$ density of states compared to that of graphene.

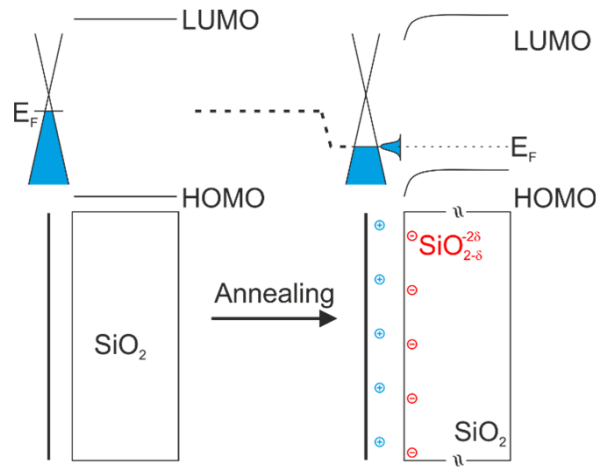

**Figure S6.** Sketch of the band structure of graphene on the $SiO_2$ substrate before and after annealing (energies are not in scale).

**Device fabrication**

To perform electrical characterization of the heterostack in a four-wire configuration, well-defined channels in the material and metal electrodes on top were required. The devices were fabricated via standard electron-beam lithography (EBL) followed by oxygen reactive ion etching (RIE) and metallization.

Fabrication steps are summarized below (see Figure S7):

a) After the transfer of single-crystal graphene arrays on $SiO_2$, the graphene heterostructure was defined by patterning with EBL the areas surrounding the desire channels. Graphene unprotected by the resist was etched by means of RIE using argon (5 sccm) and oxygen (80 sccm).

b) WS$_2$ was synthesized via CVD. The possibility of a short between contacts due to WS$_2$ is excluded, owing to the fact that WS$_2$ grows on the whole graphene crystal except in the etched areas, as carefully confirmed by SEM and Raman measurements. This represents a significant advantage in the technological process step, since otherwise WS$_2$ should have been etched with thetrafluoromethane (CF$_4$), which easily crosslinks the resist, making it impossible to dissolve.
c) Metal contacts on top of the heterostack were patterned by EBL and thermal evaporation of 60 nm of gold on top of 10 nm of titanium, followed by lift-off.

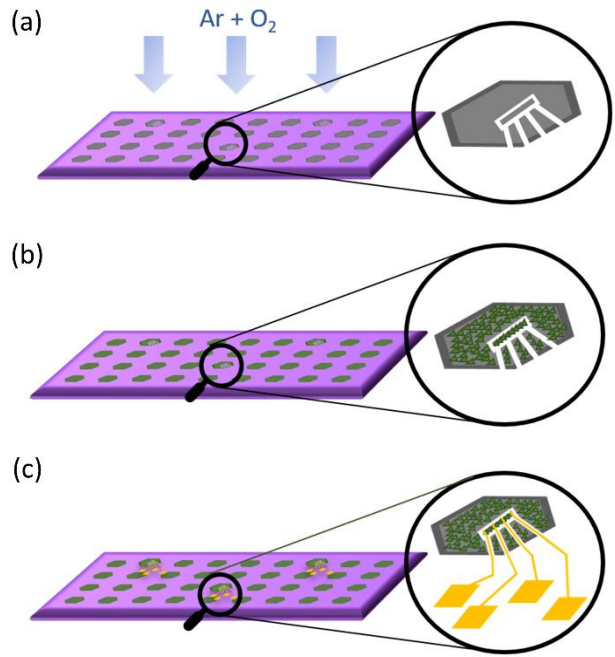

**Figure S7.** Sketch of the fabrication steps: a) graphene after etching, b) WS$_2$ growth, c) metal contacts fabrication.

**Estimation of the percentage of sulfur vacancies in WS$_2$**

In order to provide a support to the sulfur defect states proposed in the paper, we carried out XPS measurements on the WS$_2$/Gr/SiO$_2$/Si sample and recorded the W 4f and S 2p peaks with Mg K$\alpha$ and a 60° take-off angle.

At first glance, we see that the W oxide component is stronger than the disulfide component. This strong (oxidic) peak results from both the back-formation of WO$_3$ after the desorption of sulfur from the initial WS$_2$ film and from the presence of unreacted WO$_3$ remaining outside the graphene crystals, this latter being confirmed by Raman spectroscopy performed on uncovered SiO$_2$ regions (Figure S8(a)). Given the characteristics of the XPS optics indeed, the gathered signal is averaged over a significant area (about 1 mm).

The experimental evidence of unreacted WO$_3$ on SiO$_2$ allows us to formulate a likely hypothesis for the preferential growth of WS$_2$ on graphene. It is reasonable that the desorption of oxygen from the SiO$_2$ substrate hinders the sulfurization process of the WO$_3$ precursor, slowing down the WS$_2$ formation on the SiO$_2$ substrate.

The back-formation of WO$_3$ should not contribute in any way to the presence of intragap trap states.

To estimate the actual concentration of sulfur vacancies in WS$_2$, we carried out quantitative XPS analysis in the following way:

$$\frac{[W]}{[S]} = \frac{I_W}{I_S} \frac{\lambda_S}{\lambda_W} \frac{\sigma_S}{\sigma_W}$$

where $I_W$ is the integrated area of the WS$_2$ fitted component (i.e. the green area), $I_S$ is the integrated area of the S 2p full peak, $\sigma$ is the analyzer's relative sensitivity factor, which includes also the spectrometer's étendue. $\lambda$ is the effective attenuation length (EAL) at the measured kinetic energy.

The EAL for W and S were calculated utilizing the TPP-2M formula and correcting the value for the elastic scattering coefficient and the photoelectron emission angle. The ratio $\frac{\lambda_S}{\lambda_W}$ is found to be 0.91.

The relative sensitivity factors were tabulated from SPECS for the Phoibos150 analyzer with respect to F 1s and are 5.75 for the W 4f 7/2 and 1.74 for the S 2p.

Substituting the measured $\frac{I_W}{I_S}$ ratio 1.917 and the calculated values into the above formula, we find [W]/[S]=0.523, which implies the presence of about 4.5% of sulfur vacancies in the WS$_2$.

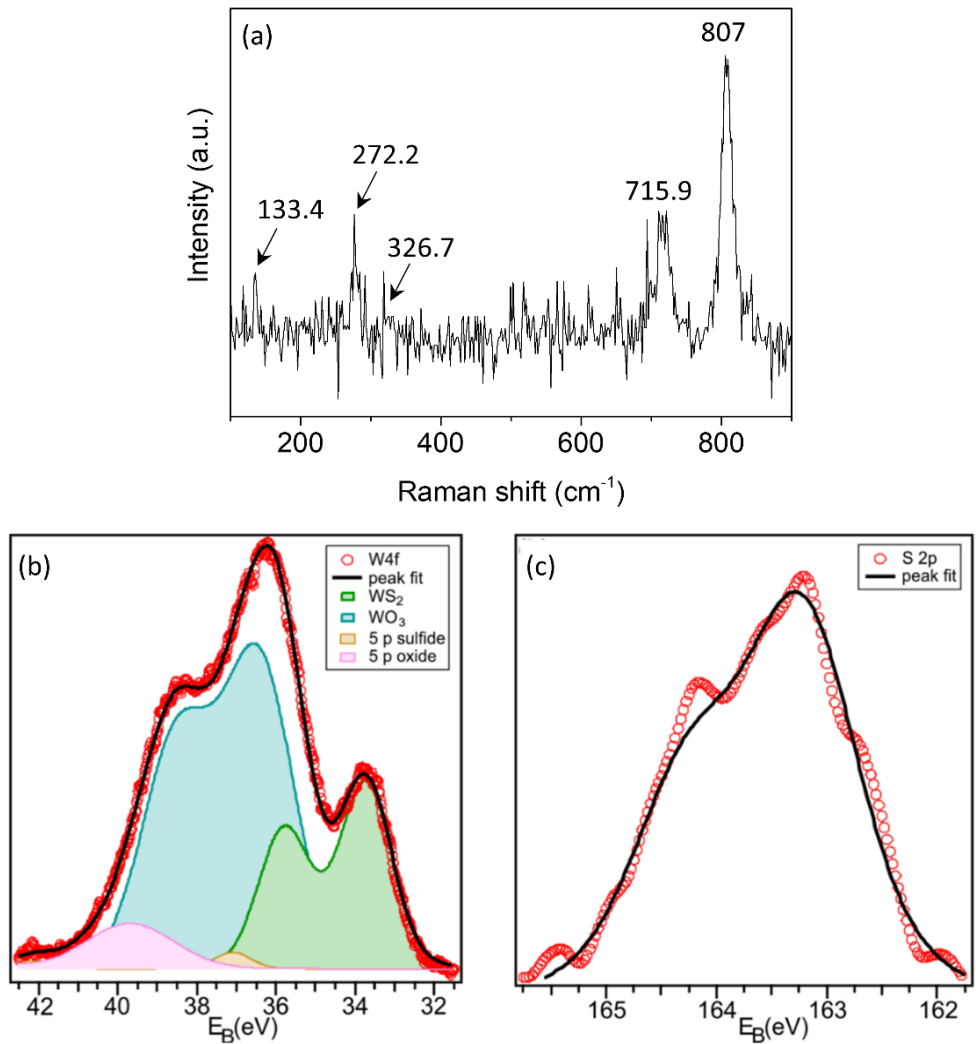

**Figure S8.** (a) Raman spectrum of $WO_3$ taken on a $SiO_2$ region not covered by the graphene/$WS_2$ heterostructure, with the characteristics peaks labeled. The signal to noise ratio is weak owing to the small amount of material. (b)-(c) XPS measurements of the W 4f (b) and S 2p (c) peaks of $WS_2$.